Trouble with Physics?

===================================================================
A review of: "The Trouble with Physics: The Rise of String Theory, the Fall of a Science, and What Comes Next", by Lee Smolin (Houghton Mifflin, Boston, N.Y. (2006))
===================================================================

Superstring Theory has been the subject of intense study by a substantial segment of the theoretical physics community for over two decades. Its goal is extremely ambitious, to say the least: nothing less than a unified quantum framework for all the fundamental interactions of matter, a "Theory of Everything" [1,2]. Its mathematical intricacies, however, are barely understandable even to a majority of physicists, and it has yet to prove itself a functioning physical theory. Nevertheless, in "The Trouble with Physics" [3], Lee Smolin tries to give an overview, including the background, history, motivation and content, as well as a detailed critique, at a level accessible to a general readership.

To help understand why a physicist might want to address an audience that cannot be expected to comprehend such a subject's hermetic details, it should be mentioned that only about a third of the book is concerned with String Theory per se. The remainder consists of an earnest plea for two things. The first is greater attention to alternative approaches to the fundamental questions of theoretical physics, in particular the line of research concerning quantum gravity that comprises the author's own main interests; the second is a critique of the assumptions and social pressures of the research community in which he works, with suggestions for improvements. Nearly as much of this book is devoted to the sociology of science - more specifically, the String Theory community - as to science itself.

In view of the exceptionally high level of mathematical preparation that a genuine understanding of the subject would require, only a selective, nontechnical description of the ingredients can be given in such a work. To make this at all meaningful is a very difficult task. Smolin succeeds in giving the interested general reader an idea of the flavour of the subject, and a vantage point from which to make some sense of the critiques that follow.

It is hard to disagree with the general socio-academic critiques, but many are far from unique to this particular field; in fact, they are rather commonplace within the broader academic community: the pressures upon young researchers to fit into an already established "niche", the lack of incentive or reward for independent thought, the tendency towards "tribalism" and "group thought", the tendency by more senior scientists, who can provide or deny opportunities to younger researchers, to judge the merits of candidates by how well they agree with their own outlook and interests, etc.

To these general sociological critiques, Smolin adds some that are more specific to the String Theory community. This is an exceptional one, both in the incomparable ambitiousness of its scientific agenda and, according to the author, its particular susceptibility to the "group thought" phenomenon. Apart from this, many would agree that String Theory suffers from a central defect that sets it apart from nearly all other scientific pursuits: it is unable as yet to provide anything that may be subjected to the test of experimental verification, beyond what is already adequately provided by more conventional frameworks, such as the "Standard Model".  From the viewpoint of the Scientific Method this is anathema, putting the subject into a very difficult position to defend. Is it really physics, or just mathematical conjecture that is too incomplete to stand up as a physical theory? All depends on promises of things to come.

The Standard Model has been very successful in accounting for observable phenomena involving the electromagnetic and Weak nuclear forces as parts of a unified theory of electroweak interactions  It also includes a consistent framework, Quantum Chromodynamics (QCD), for the Strong nuclear forces, which hold the atomic nucleus together, although these are not yet fully understood at a sufficiently detailed level to be able to account for the huge quantity of strong interaction data accumulated over decades in high energy physics laboratories. Moreover, the short range nature of the Strong interactions remains to be satisfactorily explained as a direct consequence of the model. This is all ascribed to the complicated collective effects that are necessarily present in such a relativistic many-body setting, in which any number of particles may be created or destroyed within minute time and space intervals.

The discovery of this framework and proof of its consistency were hailed as a great breakthrough in our understanding of the laws governing the interactions of elementary particles. It followed four decades of struggle, was consolidated within the short period 1970-74, and confirmed to be in agreement with experiment in the years that followed. It led within the subsequent two decades to the award of three Nobel prizes to eight theoretical physicists, and six more to experimentalists who discovered or confirmed many of its observable consequences. The theory seems at least to have no intrinsic defects other than those shared by any relativistic quantum field theoretic model.

These include, of course, one unusual feature that physicists have learned to live with for decades; namely, the fact that perturbation theory, based on successive approximations, leads at first to infinite quantities which must be eliminated through a scheme of infinite renormalization before arriving at anything that may be compared with experiment. However, this is generally not seen as an essential defect but rather a necessary feature of the perturbative approach, and the final results do agree with experiment to a very high accuracy - at least in the weak and electromagnetic case. The fact that perturbation theory is not really applicable to the Strong interactions is partially mitigated by the fact that calculations valid to all orders, using the "renormalization group" approach, demonstrate the existence of "Asymptotic Freedom"; i.e., the Strong interactions become arbitrarily weak at sufficiently short distances.

No known quantum field theoretical framework exists however that includes gravitation and is consistent with General Relativity. The incompatibility lies in the fact that when trying to treat gravitation as a quantum field, infinities persist in the perturbation theoretic calculations that cannot be eliminated through renormalization. This problem has never been overcome in any quantum field theory setting. The desire to include gravitation within a quantum framework unifying all the fundamental interactions of physics has been the main justification for the huge effort to develop String Theory over the past twenty-five years. This alternative to quantum field theory conceives of all particles, including the graviton, as quantum excitations of strings. This is very different from the usual quantum field theory framework, but the latter is expected to be recoverable in a suitable "low energy" regime (which includes all energies accessible to high energy laboratories to date).

Smolin's own research priorities are spelled out in the first chapter, entitled "The Five Great Problems in Theoretical Physics". In his view, these are: 1) To combine General Relativity and Quantum Theory in a single, complete theory of nature; 2) To resolve what he regards as problems in the foundations of Quantum Mechanics. These include finding a better epistemological explanation of quantum theory, or an alternative theory that is not based on the necessity of involving the observer as part of the measurement process; 3) To come up with a unified theory that combines all the known forces of nature (a bit redundant, when compared with Great Problem 1); 4) To explain all the apparently arbitrary constants appearing in the "Standard Model"; 5) To account for the existence (or nonexistence) of "Dark Matter" and "Dark Energy". (This is a conjectural explanation of the apparent inconsistency between the amount of observed matter in galaxies and the rates at which stars and galaxies are moving, which supposes the existence of matter and energy that is "invisible", adding up to as much as 96% of the total energy in the universe.)

These may all be very worthy goals, but not everyone would agree on their primacy,

or achievability. In fact, there are many other outstanding problems that are at least as worthy of attention, but do not appear on this list. For example, the so-called "Mass Gap Problem", which was one of the "Millennium Prize Problems" [4] announced as a challenge in the year 2000 by the Clay Institute in their list of outstanding, mainly mathematical problems (one of which - the Poincaré conjecture - has since been solved), that would gain for the successful solver, besides fame and glory, a modest prize of $1,000,000. For a physicist, the most concrete part of this problem consists of explaining, as a consequence of QCD - or some variant - why it is that the Strong nuclear interactions are short range. Although the short range nature of the Weak interactions follows directly from the Standard Model through the mechanism of "spontaneous symmetry breaking", an essential ingredient, the existence of the massive, spinless "Higgs Boson" remains to be confirmed experimentally. QCD, however, does not allow for a similar breaking of the underlying gauge symmetry governing the strong interactions. Phenomenological explanations, based on "flux tubes", and "quark confinement" (neither of which have actually been demonstrated to follow directly from the model) have been put forward. But the fact remains that the Weak and Electromagnetic interactions can be dealt with directly; i.e., the scattering amplitudes, decay rates and correlations can be computed and compared, successfully, with experiment, while the same cannot as yet be done for the Strong interactions, using just QCD as a starting point.

   No one expects to improve upon the accuracy of the results of perturbative calculations, but it would be nice to have them on a logically complete mathematical footing. Some (though not all) theoreticians see it as desirable that the foundations of quantum field theory be revamped, to make the existence of interacting quantum fields accord with our current level of mathematical understanding. (This is, roughly, the other part of the Millennium Problem referred to above.)

   These goals are no less compelling than the ones listed by Smolin, and probably several more could be added. Perhaps the reason for his particular choices was to keep within a range that could possibly be addressed by String Theory (although he concedes that 2) is not one of these). Or perhaps, the others are not listed because they consist of improving or completing our understanding of an already existing theory, and therefore do not figure as "revolutionary" (cf. T.S. Kuhn [5]).

   Although Smolin repeatedly emphasizes his preference for independent approaches, and abhorrence of "group thought", it seems that some assumptions of the community he is criticizing have also been adopted by him. In particular, little mention is made in "The Trouble with Physics" of approaches other than String Theory, except for his own specialty of Loop Gravity, which is given nearly equal prominence and space, though it only represents the interests of a very small sector of the research community.

   The main scientific critiques of String Theory elaborated in the book are:
1) The necessity for introducing unobservable "higher dimensions" that are understood as spontaneously "curled up" to such small sizes that they escape detection, without any dynamical mechanism implying such a process, or its stability, and no physical explanation of why one or another of these "backgrounds", (in ten, or eleven, dimensions) is preferred. The vast multiplicity of possible background geometries available in the "string landscape" seem to make it impossible to arrive at definite predictions since they introduce a huge number of additional parameters (the "moduli") whose values the theory is incapable of determining;
2) The lack of any experimental evidence for the distinct consequences of string theory. In particular, the essential ingredient of "supersymmetry" (from which derives the name "Superstrings") is required to assure finiteness. At the very least, this means a matched pairing of all the fermionic particles in the universe, having 1/2 integer spins (such as electrons, neutrinos, quarks) with bosonic partners, sharing similar properties, but having integer spin. Nothing remotely like this exists in the currently observed spectrum of elementary particles. The usual reply to this is: "Yes, but it is a spontaneously broken symmetry", which just suggests sweeping an essential feature under the rug, because it is inconvenient to face up to its observable implications;
3) The absence of a "complete" version of string theory. In particular, there is no "Superstring Field Theory", that would allow for processes involving the creation and annihilation of strings.

Beyond these broad critiques, there are a number of doubts raised regarding more technical points, such as the interpretation of the various types of "duality" that enter into the theory, which help to reduce what might be five distinct versions to one, and the sign of the cosmological constant, which has direct implications for the behaviour of matter in the large. Whether right or wrong, these are best left to the specialists to debate, since they do not in themselves imply fundamental trouble with the theory, simply differences of view regarding its implications. (See e.g. [6] for a discussion and rebuttal of some of these points.)
Smolin also regrets that an entire generation of theoretical elementary particle physicists has been raised to work with such a highly esoteric mathematical model, founded on far-fetched speculation, without seeing the necessity for pinning down their conjectures by precise mathematical demonstration, and without adequate consideration of possible alternatives. Finally, there are the further sociological-psychological critiques: the arrogance with which many string theorists have vaunted the relative importance of their work, and the pressure that has been put upon upcoming researchers to work along currently accepted lines, at risk of achieving nothing of any real significance (and getting no employment). The criticism of immodest swagger however comes from an author who ends his book by stating, with no hint of irony or self-mockery, that he will now return to his "real job", which is to "finish the revolution that Einstein started".
Although some of the most brilliant and talented researchers of our day have devoted themselves to String Theory, it is undeniable that advances have largely been driven by far-reaching, but incompletely demonstrated conjectures. New ideas have had an esoteric, self-contained character, independent of the traditional checks based on careful comparison with experiment. There is no doubt that Einstein, who always was concerned with testing theoretical predictions against observation, would be astonished at such a situation persisting after more than two decades of research in the subject. The most important concern for String Theory, in the end, may be that the legacy of the current generation of theorists risks being purely mathematical in nature if no contact is made within a reasonable time with experimentally verifiable phenomena.

Whatever the validity of the arguments in this book as a critique of String Theory, they do not justify the sweeping conclusion implied in the title. "The Trouble with String Theory" might have sold fewer copies, but it would have been a more accurate characterization of the content of this book. It is a poor service to potential young physicists, and an injustice to the rest of physics, to dismiss as unworthy of mention under the title "The Trouble with Physics" the many wonderful discoveries, both experimental and theoretical, that have been made over the past three decades in other areas of physics.
The most interesting of these, which have merited not a few Nobel Prizes in recent years, have tended to concern the "other" frontier of experimental and theoretical physics: the physics of temperatures very near to absolute zero and macroscopic quantum phenomena. The latter largely concern quantum effects and states of matter that can be observed at or near macroscopic scales, both at very low temperatures, and at temperatures at which it was never previously imagined such collective quantum effects could occur. These include many of the really remarkable achievements of physics of the past few decades. For example, there was the creation, in plasma physics laboratories, of a "Bose-Einstein Condensate", an entirely new state of matter, achievable only at the lowest temperatures, predicted by Einstein in the 1920's, but not observed in a laboratory until the 1990's [7]. Related to this is the remarkable use of "laser cooling" and optical or other methods for reducing the motion of individual molecules to such a slow point that they can be individually studied and controlled, and together with this, the creation of purely optical grids, and traps [8]. There have also been important theoretical advances in the understanding of disordered systems, liquid crystals and polymer dynamics [9]. There have been the observation and theoretical explanation of the "fractional quantum Hall effect" a phenomenon that brings to light the beautiful role played by topology in the electro-magnetic properties of ordinary matter in a suitable quantum regime [10]. The discovery of high temperature superconductivity has led to fundamental challenges to explain associated recently discovered phenomena. Another dramatic development in theoretical physics has concerned the precise understanding of scaling properties in critical phenomena in 2-dimensions, and related phenomena such as percolation, and critical wetting. In the high energy domain it should also be mentioned that the prediction of "asymptotic freedom" [11], which was an essential step in making sense of a quantum field theory of the strong interactions, obtained its experimental verification within the past two decades.

   There are numerous other examples that could be cited. Perhaps none of the above can be compared in ambition or scope with seeking a "Theory of Everything", nor to the impact of the revolutionary developments of the early twentieth century: Relativity Theory and Quantum Mechanics. They nevertheless represent very exciting advances in our understanding of the physical world, and provide abundant evidence that much more is healthy and robust in the physics of the past three decades than what might, by narrowly focusing on the shortfalls of String Theory, lead to the conclusion that Physics of our time is in trouble.

John Harnad
Centre de recherches mathématiques
Université de Montréal
C.P.6127 Succ. centre ville
Montreal, Québec H3C 3J7